\documentstyle[12pt]{article}
\begin{document}

\title{Scheme for the implementation of a universal quantum cloning
machine via cavity-assisted atomic collisions in cavity QED }
\author{XuBo Zou, K. Pahlke and W. Mathis\\
Electromagnetic Theory Group at THT,\\
 Department of Electrical
Engineering, \\
University of Hannover, Germany}

\date{\today}
\maketitle
\begin{abstract}
We propose a scheme to implement the $1\rightarrow2$ universal
quantum cloning machine of Buzek et.al [Phys. Rev.A 54,
1844(1996)] in the context of cavity QED. The scheme requires
cavity-assisted collision processes between atoms, which cross
through nonresonant cavity fields in the vacuum states. The cavity
fields are only virtually excited to face the decoherence problem.
That's why the requirements on the cavity quality factor can be
loosened.

\end{abstract}

In the last decade, considerable progress in the field of quantum
information processing has been made. New prospects in computation
and communication technology are very challenging. Basic questions
on this kind of information transfer have been raised. Quantum
information differs from classical information in a fundamental
way. For instance, it is not possible to construct a device that
produces an exact copy of the state of a simple quantum system
\cite{ww}. This statement is a consequence of the linearity of
quantum mechanics. It constitutes one of the most significant
differences between classical information and quantum information.
The seminal paper of Buzek and Hillery \cite{bm} put a strong
impulse on quantum cloning. This problem was extensively studied
in the example of discrete quantum variable systems, such as
quantum qubits \cite{nd} or d-level systems \cite{wbc}. Bounds on
the maximum possible fidelity of the clones produced by universal
quantum cloning machine was derived and an optimal universal
quantum cloning transformation was discovered \cite{nd,wbc}.\\
In order to make new application in this field possible, an
appropriate quantum system is needed, which can be very well
isolated from the environment to suppress decoherence processes.
Several physical systems were suggested to implement the concept
of quantum information processing: cavity QED
 \cite{tc}, trapped ion systems \cite{cd} and nuclear magnetic
resonance systems \cite{cg}. Cavity QED with Rydberg atoms, which
cross superconducting cavities, are nearly ideal systems for this
purpose. Various entangled states such as EPR pairs \cite{two} and
GHZ states \cite{three} have been successfully produced by a
successive interaction of a series of atoms with the cavity field.
An experimental implementation of the quantum logic gate
\cite{logic} and the absorption-free detection of a single photon
\cite{mea} have been reported by using a resonant atom-cavity
interaction. A number of schemes have been proposed for the
teleportation of atomic states \cite{cavity}, the implementation
of quantum algorithms \cite{scu,qu} and the realization of
entanglement purification \cite{filter1}. Recently, quantum
cloning of a single photon state was demonstrated experimentally
\cite{clon1} by using the scheme, which was proposed in Ref
\cite{clon2}. An alternative experimental implementation of the
cloning network, which is based on the NMR system, has been
reported \cite{clon3}. More recently, a cavity QED scheme is
proposed to implement a $1\rightarrow2$ universal quantum cloning
machine by using a resonant atom-cavity interaction \cite{clon4}.
In this scheme, cavities act as memories, which store the
information of an electric system and transfer it back to this
electric system after the conditional dynamics. But, the
decoherence of the cavity field becomes one of the main obstacles
for the experimental realization of the universal quantum cloning
machine.

In this paper, we propose scheme to implement the $1\rightarrow2$
universal quantum cloning machine of Buzek et al \cite{bm} within
cavity QED. In contrast to the scheme \cite{clon4}, our scheme
requires a cavity-assisted collisions between atoms \cite{zheng}.
This technique has been experimentally demonstrated \cite{ppa}. In
order to overcome the main problem of decoherence a virtually
excitation of the cavity field is chosen. That's why this kind of
implementation becomes essentially insensitive to cavity losses
and to thermal cavity excitations. The cavity-assisted collision
processes have been used for the generation of entangled atomic
state \cite{state} and
the implementation of quantum search algorithm \cite{qu}.\\

At first, we consider the interaction of $N$ two-level atoms with
a single-mode cavity field. In the interaction picture the
Hamiltonian is
\begin{eqnarray}
H=g\sum_{j=1}^N(e^{-i\delta{t}}a^{\dagger}\sigma_{-}^j+e^{i\delta{t}}a\sigma_{+}^j)\,,
\label{1}
\end{eqnarray} where $\sigma_{-}^j=|g_j\rangle\langle{e_j}|$
and $\sigma_{+}^j=|e_j\rangle\langle{g_j}|$, with $|g_j\rangle$
and $|e_j\rangle$ ($j=1,\cdots,N$) are the ground states and the
excited states of the $j$th atom. The annihilation and creation
operator of the cavity field are $a$ and $a^{\dagger}$. We use $g$
 as the atom-cavity coupling strength and $\delta$ as the detuning
between the atomic transition frequency and the cavity frequency.
We consider the case $\delta\gg{g\sqrt{\bar{n}+1}}$, with the mean
photon number $\bar{n}$ of the cavity field. With this restriction
it is convenient to consider the interaction (\ref{1}) in terms of
a coarse-grained Hamiltonian, which neglects the effect of rapidly
oscillating terms. We use the time-averaging method of Ref
\cite{james} to derive at the effective Hamiltonian
\cite{zheng,state}
\begin{eqnarray}
H=\lambda[\sum_{j=1}^N(|e_j\rangle\langle
e_j|aa^{\dagger}-|g_j\rangle\langle
g_j|a^{\dagger}a)+\sum_{j,k=1}^N\sigma_{+}^j\sigma_{-}^k], ~~j\neq
k \,,\label{2}
\end{eqnarray}
with the abbreviation $\lambda=g^2/\delta$. If the cavity field is
initially in the vacuum state, the effective Hamiltonian reduces
to
\begin{eqnarray}
H=\lambda(\sum_{j=1}^N|e_j\rangle\langle
e_j|+\sum_{j,k=1}^N\sigma_{+}^j\sigma_{-}^k), ~~j\neq k\,.
\label{3}
\end{eqnarray}
Now we present the scheme to implement the $1\rightarrow2$
universal quantum cloning machine of Buzek et al \cite{bm}. The
experimental setup, which we suggest to use, is depicted in Fig.1.
Three two-level atoms are needed to implement our scheme.
Furthermore, we assume that all cavities are prepared in the
vacuum state. The atoms interact with each other via
cavity-assisted atomic collision processes (\ref{3}). We assume
that atom $1$ carries the quantum state, which will be cloned. It
is prepared in an arbitrary pure state of the form
\begin{eqnarray}
|\Psi_{in}\rangle=\alpha|g_1\rangle+\beta|e_1\rangle\,, \label{4}
\end{eqnarray}
where $\alpha$ and $\beta$ are complex coefficients. The other
atoms $2$ and $3$ are prepared in the state
$|e_2\rangle|g_3\rangle$. They are sent into the first cavity $A$.
The effective interaction (\ref{3}) between the atoms $2$ and $3$
can be written as
\begin{eqnarray}
H=\lambda(|e_2\rangle\langle{e_2}|+|e_3\rangle\langle{e_3}|+\sigma_{+}^2\sigma_{-}^3+\sigma_{-}^2\sigma_{+}^3)
\,.\label{5}
\end{eqnarray}
The evolution of the quantum state of the atoms $2$ and $3$ is
given by
\begin{eqnarray}
|e_2\rangle|g_3\rangle\rightarrow
e^{-i\lambda{t}}[\cos(\lambda{t})|e_2\rangle|g_3\rangle-i\sin(\lambda{t})|g_2\rangle|e_3\rangle]
\,.\label{6}
\end{eqnarray}
If we choose the interaction time to satisfy $\lambda{t}=\pi/4$,
the state of the system becomes
\begin{eqnarray}
|\Psi_{23}\rangle=\frac{1}{\sqrt{2}}[|e_2\rangle|g_3\rangle-i|g_2\rangle|e_3\rangle]
\,,\label{7}
\end{eqnarray}
where we have discarded the common phase factor. After the atoms
$2$ and $3$ emit from the cavity $A$, these two atoms and the atom
$1$, which carries the information, will simultaneously be sent to
another cavity $B$. But before, a classical field is applied to
rotate atom $1$ and atom $2$ along the $z$-axis by the angles
$\theta_1$ and $\theta_2$, respectively. The corresponding
transformation is given by
\begin{eqnarray}
U_j=\exp[-i\theta_j(|e_j\rangle\langle{e_j}|-|g_j\rangle\langle{g_j}|)]\,.
\label{8}
\end{eqnarray}
Thus, the state $|\Psi_{in}\rangle$ becomes
\begin{eqnarray}
|\Psi_{in}^{\prime}\rangle=\alpha{e^{i\theta_1}}|g_1\rangle+\beta{e^{-i\theta_1}}|e_1\rangle
\,,\label{9}
\end{eqnarray}
where $\theta_1$ will be determined later. If we choose the
parameter $\theta_2=\pi/4$, the two-atom quantum state
$|\Psi_{23}\rangle$ becomes
\begin{eqnarray}
|\Psi_{+}\rangle=\frac{1}{\sqrt{2}}[|e_2\rangle|g_3\rangle+|g_2\rangle|e_3\rangle]
\,.\label{10}
\end{eqnarray}
The state of the total system
$|\Psi_{in}^{\prime}\rangle\otimes|\Psi_{+}\rangle$ can directly
be used to implement the $1\rightarrow2$ universal quantum cloning
machine. Now we send three atoms into the cavity $B$. These three
atoms interact with each other via cavity-assisted atomic
collision processes according to the effective Hamiltonian
(\ref{3}):
\begin{eqnarray}
H=\lambda(\sum_{j=1}^3|e_j\rangle\langle{e_j}|+\sigma_{+}^1\sigma_{-}^2+\sigma_{-}^1\sigma_{+}^2+
\sigma_{+}^1\sigma_{-}^3+\sigma_{-}^1\sigma_{+}^3+\sigma_{+}^2\sigma_{-}^3+\sigma_{-}^2\sigma_{+}^3)
\,.\label{11}
\end{eqnarray}
If the atoms are in the state $|g_1\rangle|\Psi_{+}\rangle$ or
$|e_1\rangle|\Psi_{+}\rangle$, the quantum state evolves as
follows:
\begin{eqnarray}
|g_1\rangle|\Psi_{+}\rangle &\rightarrow& e^{-i3\lambda{t}/2}
\{[\cos(\frac{3}{2}\lambda{t})
-\frac{i}{3}\sin(\frac{3}{2}\lambda{t})]|g_1\rangle|\Psi_{+}\rangle\nonumber\\
&&-\frac{i2\sqrt{2}}{3}\sin(\frac{3}{2}\lambda{t})|e_1\rangle|g_2\rangle|g_3\rangle
\}\nonumber\\
|e_1\rangle|\Psi_{+}\rangle &\rightarrow& e^{-i5\lambda{t}/2}
\{[\cos(\frac{3}{2}\lambda{t})
-\frac{i}{3}\sin(\frac{3}{2}\lambda{t})]|e_1\rangle|\Psi_{+}\rangle\nonumber\\
&&-\frac{i2\sqrt{2}}{3}\sin(\frac{3}{2}\lambda{t})|g_1\rangle|e_2\rangle|e_3\rangle\}
\,.\label{12}
\end{eqnarray}
If we choose the interaction time to satisfy $\lambda{t}=2\pi/9$,
the state of the total system
$|\Psi_{c}^{\prime}\rangle\otimes|\Psi_{+}\rangle$ evolves into
\begin{eqnarray}
|\Psi_m\rangle&=&\alpha{e^{i\theta_1+i\pi/6}}\left[\sqrt{\frac{2}{3}}|e_1\rangle|g_2\rangle|g_3\rangle+
\sqrt{\frac{1}{3}}e^{i\pi/3}|g_1\rangle|\Psi_+\rangle\right]\nonumber\\
&&+\beta{e^{-i\theta_1+i\pi/18}}\left[\sqrt{\frac{2}{3}}|g_1\rangle|e_2\rangle|e_3\rangle+
\sqrt{\frac{1}{3}}e^{i\pi/3}|e_1\rangle|\Psi_+\rangle\right]\,.
\label{13}
\end{eqnarray}
After the three atoms have crossed through the cavity $B$, we
again send the atoms $2$ and $3$ into the third cavity $C$. The
effective interaction between the atoms can be described by
Eq.(\ref{5}). After the interaction time $\tau$, the quantum state
(\ref{13}) evolves into
\begin{eqnarray}
|\Psi_n\rangle&=&\alpha{e^{i\theta_1+i\pi/6}}[\sqrt{\frac{2}{3}}|e_1\rangle|g_2\rangle|g_3\rangle+
\sqrt{\frac{1}{3}}e^{i\pi/3-2i\lambda\tau}|g_1\rangle|\Psi_+\rangle]
\nonumber\\
&&+\beta{e^{-i\theta_1+i\pi/18}}[\sqrt{\frac{2}{3}}e^{-2i\lambda\tau}|g_1\rangle|e_2\rangle|e_3\rangle+
\sqrt{\frac{1}{3}}e^{i\pi/3-2i\lambda\tau}|e_1\rangle|\Psi_+\rangle]\,.
\label{14}
\end{eqnarray}
We will determine the interaction time $\tau$ later. After the
atom $2$ and the atom $3$ have crossed through the cavity $C$,
these two atoms are rotated along the $z$-axis by the angles
$\theta_3$ and $\theta_4$. Therefore a classical microwave pulses
is used. The corresponding transformation is described by
Eq.(\ref{8}). If we choose the condition $\theta_3=\theta_4$, the
quantum state (\ref{14}) becomes
\begin{eqnarray}
|\Psi_o\rangle&=&\alpha{e^{i\theta_1+i\pi/6}}[\sqrt{\frac{2}{3}}
e^{-2i\theta_3}|e_1\rangle|g_2\rangle|g_3\rangle+
\sqrt{\frac{1}{3}}e^{i\pi/3-2i\lambda\tau}|g_1\rangle|\Psi_+\rangle]\nonumber\\
&&+\beta{e^{-i\theta_1+i\pi/18}}[\sqrt{\frac{2}{3}}e^{-2i\lambda\tau+2i\theta_3}
|g_1\rangle|e_2\rangle|e_3\rangle+
\sqrt{\frac{1}{3}}e^{i\pi/3-2i\lambda\tau}|e_1\rangle|\Psi_+\rangle]\,.
\label{15}
\end{eqnarray}
We choose the parameters $\theta_1$, $\theta_3$ and $\lambda\tau$
to satisfy $\theta_1=-\pi/18$, $\theta_3=\pi/6$ and
$\lambda\tau=\pi/3$. In this case the state (\ref{15}) reduces to
\begin{eqnarray}
|\Psi_f\rangle&=&\alpha[\sqrt{\frac{2}{3}}|e_1\rangle|g_2\rangle|g_3\rangle+
\sqrt{\frac{1}{3}}|g_1\rangle|\Psi_+\rangle]\nonumber\\
&&+\beta[\sqrt{\frac{2}{3}}|g_1\rangle|e_2\rangle|e_3\rangle+
\sqrt{\frac{1}{3}}|e_1\rangle|\Psi_+\rangle] \,,\label{16}
\end{eqnarray}
if the common phase factor is discarded. This equation
demonstrates, that the optimal $1\rightarrow2$ cloning process is
implemented \cite{bm}.\\

In summary, a scheme for the implementation of the optimal
$1\rightarrow2$ cloning process is proposed. In contrast to the
scheme \cite{clon4}, only simple two-level atoms are required,
which interact with the cavity fields. This might simplify the
experimental implementation of the scheme of Buzek et al. In
contrast to the scheme of Milman et al \cite{clon4}, our scheme
requires a cavity-assisted collisions between atoms \cite{zheng}.
This technique has been experimentally demonstrated \cite{ppa}. In
this scheme, the cavity field is only virtually excited. There is
no transfer of quantum information between atoms and cavity
fields. That's why the requirement on the quality factor of the
cavity can be greatly loosened. This scheme is essentially
insensitive to cavity losses and to thermal cavity excitations. It
should be pointed out that the presented scheme involves three
cavities, which might make the experimental implementation of the
present scheme more complicated.\\
Finally we give a brief discussion on the experimental feasibility
of the presented scheme. To implement the scheme, we need to
preserve the coherence of the cavity field before the atoms are
flying out of the cavity. For the Rydberg atoms with principle
quantum number $50$ and $51$, the radiative time is about
$T_r=3\times10^{-2}$seconds. The coupling of the atoms and the
cavity field is $g/2\pi=50$kHz \cite{ppa}. In order to control the
entanglement in the cavity-assisted collision process, the
detuning $\delta$ should be much greater than $g$. With the choice
$\delta=10g$ the interaction time between the atom and the cavity
field is of the order $\pi\delta/g^2=10^{-4}$seconds. At this
scale the time, which is needed to rotate the single qubit, is
negligible. Thus, the interaction time, which is needed to perform
the total procedure is shorter than the time, which is needed by
the scheme \cite{clon4}. This time interval is much shorter than
$T_r$ and the photon lifetime $1ms$ in the present cavity.
Therefore, based on the cavity QED technique the presented scheme
is realizable.

\begin{flushleft}

{\Large \bf Figure Captions}

\vspace{\baselineskip}

{\bf Figure1.} This is the schematic diagram of the optimal
$1\rightarrow2$ quantum cloning process, which copies the quantum
state of the atom $1$ to the atoms $2$ and $3$. Three cavities
$A$, $B$ and $C$ are involved, which are prepared in the vacuum
state. The abbreviations $R_i$ denote the Ramsey zones, in which a
classical field rotates the atoms along the z axis by $\theta_i$.
At first the atoms $2$ and $3$ enter the cavity $A$, where they
are prepared in a maximally entangled state. After the atoms $1$
and $2$ have been manipulated by classical fields, the three atoms
are simultaneously sent into the cavity $B$, where they interact
with each other via cavity-assisted atomic collision processes.
After the atoms $2$ and $3$ have crossed through the cavity $B$,
the cavity $C$ and two classical fields are used to perform phase
shift operations on the quantum states of the atoms $2$ and $3$.

\end{flushleft}


\begin{thebibliography}{99}
\bibitem{ww}W. K. Wootters and W. H. Zurek, Nature (London)299, 802 (1982).
\bibitem{bm}V. Buek and M. Hillery, Phys. Rev. A54, 1844 (1996).
\bibitem{nd}N. Gisin and S. Massar, Phys. Rev. Lett.79, 2153 (1997);
D. Bruss, D. P. DiVincenzo, A. Ekert, C. A. Fuchs, C.
Macchiavello, and J. A. Smolin, Phys. Rev. A57, 2368 (1998). C.-S.
Niu and R. B. Griffiths, Phys. Rev. A58, 4377 (1998). N. J. Cerf,
Phys. Rev. Lett.84, 4497 (2000). D. Bruss, A. Ekert, and C.
Macchiavello, Phys. Rev. Lett.81, 2598 (1998);
\bibitem{wbc}R. F. Werner, Phys. Rev. A 58, 1827 (1998).
V. Buek and M. Hillery, Phys. Rev. Lett.81, 5003 (1998). N. J.
Cerf, J. Mod. Opt.47, 187 (2000).
\bibitem{tc}Q. A. Turchette, C. J. Hood, W. Lange, H. Mabuchi and H. J. Kimble
Phys. Rev. Lett. 75(1995)4710
\bibitem{cd}C. Monroe, D. M. Meekhof, B. E. King, W.M. Itano and D.
J. Wineland, Phys. Rev. Lett. 75(1995)4714, J. I. Cirac and P.
Zoller, Phys. Rev. Lett. 74(1995)4091.
\bibitem{cg}I. Chuang, N. Gershenfeld and M. Kubinec Phys. Rev. Lett. 80(1998)3408.
\bibitem{two}E. Hagley, X. Maitre, G. Nogues, C. Wunderlich, M. Brune, J. M. Raimond, and S. Haroche, Phys. Rev. Lett. 79, 1(1997).
\bibitem{three}A. Rauschenbeutel, G. Nogues, S. Osnaghi, P. Bertet, M. Brune, J. M. Raimond, and S. Haroche, Science 288,
2024(2000)
\bibitem{logic}A. Rauschenbeutel, G. Nogues, S. Osnaghi, P. Bertet, M. Brune, J. M. Raimond, and S. Haroche , Phys. Rev. Lett. 83, 5166 (1999).
\bibitem{mea}Nogues, G., A. Rauschenbeutel, S. Osnaghi, M. Brune, J. M. Raimond, and S. Haroche, 1999, Nature (London) 400, 239.
\bibitem{cavity}J.I. Cirac and A.S. Parkins,Phys. Rev. A 50, R4441 (1994)
\bibitem{scu}Marlan O. Scully and M. Suhail Zubairy, Phys. Rev. A 65, 052324
(2002); M. O. Scully and M. S. Zubairy, Proc. Natl. Acad. Sci.
U.S.A. 98, 9490 (2001).
\bibitem{qu}F. Yamaguchi, P. Milman, M. Brune, J. M. Raimond, S.
Haroche, quant-ph/0203146
\bibitem{filter1}J. L. Romero, L. Roa, J. C. Retamal, and C. Saavedra,Phys. Rev. A 65, 052319 (2002)
\bibitem{clon1} Antia Lamas-Linares, Christoph Simon, John C. Howell, Dik
Bouwmeester, Science 296, 712 (2002)
\bibitem{clon2}C. Simon, G. Weihs, and A. Zeilinger, Phys. Rev. Lett.{\bf 84}, 2993 (2000);
J. Kempe, C. Simon, and G. Weihs, Phys. Rev. A {\bf 62}, 032302
(2000)
\bibitem{clon3}H. K. Cummins, C. Jones, A. Furze, N. F. Soffe, M. Mosca, J. M. Peach, and J. A. Jones
Phys. Rev. Lett. 88, 187901 (2002)
\bibitem{clon4} P. Milman, H. Ollivier, J. M. Raimond, quant-ph/0207039
\bibitem{zheng}S. B. Zheng and G. C. Guo, Phys. Rev. Lett. 85, 2392 (2000).
\bibitem{ppa}S. Osnaghi et al., Phys. Rev. Lett. 87, 037902 (2001)
\bibitem{state}Shi-Biao Zheng, Phys. Rev. Lett. 87, 230404 (2001)
,Guo-Ping Guo et al., Phys. Rev. A 65, 042102 (2002).
\bibitem{james}D. F. James, Fortschr. Phys. 48, 823(2000)
\end{thebibliography}
\end{document}